\documentstyle [12pt]{article}
\normalbaselines
\def\baselinestretch{1.4}
\def\be{\begin{equation}}
\def\te{\end{equation}}
\def\bea{\begin{eqnarray}}
\def\nn{\nonumber}
\def\tea{\end{eqnarray}}
\def\endignore{}
\def\beginignore#1\endignore{}

\def\d{\delta}

\def\h{\eta}

\def\l{\lambda}

\newskip\humongous \humongous=0pt plus 1000pt minus 1000pt

\newif\ifdtup

\newcommand{\Ubar}{\overline U}
\newcommand{\Dbar}{\overline D}
\newcommand{\Ebar}{\overline E}

\begin{document}

\hspace{5in}{\bf UMD-PP99-052}

\begin{center}
\Large \bf WITH NEUTRINO MASSES REVEALED, PROTON DECAY IS THE MISSING
LINK \footnote{Based in part on a talk delivered at the Abdus Salam
Memorial Meeting, ICTP, Trieste, November 1997.}
\end{center}

\vspace{.5cm}

\begin{center}
Jogesh C. Pati\\ [5mm]

Physics Department, University of Maryland, College Park,\\
 MD 20742, USA\\
October, 1998\\
\vspace{.5in}

ABSTRACT
\end{center}

By way of paying tribute to Abdus Salam, I recall the ideas of higher
unification that he and I initiated. I discuss the current status of
those ideas in the light of recent developments, including those of:
(a) gauge coupling unification, (b) discovery of neutrino-oscillation at
SuperKamiokande, and (c) ongoing searches for proton decay.  It is remarked
that the mass of $\nu_\tau$ ($\sim$ 1/20 eV) suggested by the SuperK
result seems to provide clear support for an underlying unity of forces 
based
on the ideas of (i) SU(4)-color, (ii) left-right symmetry and (iii)
supersymmetry.  The change in perspective, pertaining to both gauge
coupling unification and proton decay, brought forth by supersymmetry
and superstrings is presented.  The beneficial roles of string-
symmetries in addressing certain naturalness problems of supersymmetry,
including that of rapid proton decay, are noted.  
In the last section, attention is drawn to the recent joint works with K.
Babu and F. Wilczek, where the influence of neutrino masses and thus of the
new SuperK result on proton decay are noted.
In this context, 
it is remarked that with neutrino masses and coupling
unification revealed, the discovery of 
proton decay, that remains as the missing link, should not be far behind.

\renewcommand{\baselinestretch}{1.4}

\newpage

\section*{\bf {I. Salam in Perspective}}

Abdus Salam was a great scientist and a humanitarian.  His death was indeed 
a loss to science and especially to the growth of science in the
third world.  He will surely be remembered for his contributions to
physics, some of which have proven to be of lasting value.
These include his pioneering work on electroweak unification for which he
shared the Nobel Prize in physics in 1979 with Sheldon Glashow and
Steven Weinberg.  Contribution of this calibre is rare.

But I believe his most valuable contribution to science and humanity,
one that is perhaps unparalleled in the world, is the sacrifice he has
made of his time, energy and personal comfort in promoting the cause of
science in different corners of the globe, in particular the third
world.  His lifelong efforts in this direction have led to the creation
of some outstanding research centres, including especially the
International Centre for Theoretical Physics (ICTP) at Trieste, Italy,
\footnote{Now named (at this meeting) the Abdus Salam International
Centre for Theoretical Physics.} an International Centre for Genetic 
Engineering and Biotechnology with components in Trieste and Delhi, and
an International Centre for Science and High Technology in Trieste.
Salam 
dreamed 
of creating twenty international centres like the ICTP, spread
throughout the world, emphasizing different areas of science and
technology.  
Approaching developed as well as developing nations, for
funding of such institutions, 
Salam often used the phrase: "science is not
cheap, but expenditures on it will repay tenfold" \cite{SIR}.  If only 
Salam had lived a few more years in good health, many more such
institutions would have surely come to fruition.

Salam was also a
strong supporter of world peace, and thus of nuclear disarmament and
Pugwash.  Thus, in addition to his numerous awards for his contributions
to physics, including the Nobel prize, he also received, 
some major awards for his contributions to peace and international
collaboration, including the Atoms for Peace Award in 1968 and the
"Ettore Majorana" - Science for Peace prize in 1989. It is hard to believe 
that a single individual can accomplish so much in one lifetime.  
In this sense, Salam was indeed a 
rare individual---a phenomenon.

I was especially fortunate to have collaborated with Salam closely for
over a decade.  Of this period, I treasure most the memory of many
moments which were marked by the struggle and the joy of research that
we both shared.  Needless to say, Salam played a central role in the
growth of the ideas which we initiated.  
Let me touch upon one aspect of Salam's personality.  During the ten year
period of our collaboration, there have been many letters, faxes,
arguments over the phone and in person and even heated exchanges, about
tastes and judgements in physics, but always in a good natured
spirit \footnote{A brief account of how our collaboration evolved in the
initial phase is given in my article in the Proceedings of the
Salamfestschrift \cite{SF} which was held here at ICTP in 1993 (that
is probably the last scientific meeting that Salam attended), and a
shorter version is given in the article written in his honor after he
passed away \cite{Pati}.  The first section of this talk is based in part 
on these two articles.}
In our discussions, Salam had some favorite phrases.  For example, he
would sometimes come up with an idea and get excited.  If I expressed
that I did not like it for such and such reason, he would get impatient
and say to me: "My dear sir, what do you want: Blood?" I would sometimes
reply by saying: "No Professor Salam, I would like something better".  
Whether I was
right or wrong, he never took it ill.  It is this attitude on his part
that led to a healthy collaboration and a strong bond between us.  Most 
important
for me, by providing strong encouragement from the beginning, yet often
arguing, he could bring out the best in a collaborator.  For
this I will remain grateful to him.

By way of paying tribute to Salam therefore, I would first like to
recall briefly the ideas on higher unification
which we initiated (Sec. 2), and then present their current 
status in the context of subsequent experimental and 
theoretical developments (Secs. 3,4, and 5).  
The experiments of special relevance are: (a) recent observation of 
neutrino oscillation at 
SuperKamiokande, (b) the precision measurements of 
the gauge couplings at LEP, and (c) ongoing searches for proton
decay.

In Section 3, I discuss how the recent discovery of
atmospheric neutrino-oscillation at SuperKamiokande \footnote{While the
SuperKamiokande discovery occurred some six months after the
presentation of this talk, its implications are 
included here because they are
so directly relevant
to the unification ideas
proposed in the early 70's.}, especially the mass of $\nu_\tau$
suggested by the SuperK result, on the one hand 
agrees well with the gauge coupling 
unification revealed by the extrapolation of the LEP data, 
and on the other hand provides clear support for 
the route to higher unification based on the 
concepts of SU(4)-color and left-right
symmetry.  On the theoretical side, the major 
developments of
the last two decades are the ideas of supersymmetry, superstrings,
and now M-theory.  I will briefly remark how these later developments
fully retain the basic ideas of higher unification of the 70's, and at
the same time, provide a substantially new perspective because they unify
 gravity with the other three forces (Sec.4).  
The change in perspective pertains to both gauge coupling unification
and proton decay.  
In discussing the puzzle of proton-longevity in supersymmetry, I remark
, following recent work, how string-derived
symmetries play an essential role in providing a natural resolution of this 
puzzle (Sec. 5).  In the last section, I present the results of two recent
papers by  
Babu, Wilczek and me, that exhibit a strong link between neutrino
masses and proton decay in the context of supersymmetric unification
(Sec. 6). 
Based on these works, 
I remark that the
observation of coupling unification as well as the discovery of 
neutrino-oscillation 
at SuperK strengthen our expectations for discovery of proton decay in 
the near future. 

\renewcommand{\baselinestretch}{1}        
\section*{\bf	II.	Status of Particle Physics in 1972: 
The Growth of New Ideas}
\renewcommand{\baselinestretch}{1.4}

{\large \bf IIA.} The collaborative research of Salam and 
myself started during my short
visit to Trieste in the summer of 1972.  At this time, 
the electroweak SU(2) 
$\times$ U(1)-theory existed \cite{SU(2)}, but there was no clear idea of 
the origin of the fundamental strong interaction.  The latter was
thought to be generated, for example, by the vector bosons
($\rho, \omega, K^* $ and $\phi$), or even the spin-o mesons ($\pi,
K, \eta, \eta', 
\sigma$), assumed to be elementary, 
or a neutral U(1) vector gluon coupled
universally to all the quarks \cite{GM}.  Even the existence of the
SU(3)-color degree of freedom \cite{OWG,HanNam} as a global symmetry 
was not commonly accepted, because
many thought that this would require an undue proliferation of elementary
entities.  And, of course, asymptotic freedom had not yet been
discovered.

In the context of this background, the SU(2) $\times$ U(1) theory itself
appeared (to us) as grossly incomplete, even in its gauge-sector (not to
mention the Higgs sector), because it possessed a set of scattered
multiplets, involving quark and lepton fields, 
with rather peculiar assignment of their weak hypercharge
quantum numbers.  To remove these shortcomings, we wished: (a) 
to find 
a higher symmetry-structure that would
organize the scattered multiplets together, and explain 
the seemingly arbitrary assignment of their weak hypercharges; (b) 
to provide a rationale for the co-existence of quarks and
leptons; further (c) to find a reason for the 
existence of the weak, electromagnetic as well as
strong interactions, by generating the three forces together by a
unifying gauge principle; and finally (d) to understand the 
quantization of
electric charge, regardless of the choice of the multiplets, in a way which
should also explain \footnote{We thought that if one could understand 
why the electron and the proton have equal and opposite charges, one
would have an answer to Feynman's question as to why it is that the
electron and the proton - rather than the positron and proton - exhibit
the same sign of longitudinal polarization in $\beta$-decay.  The V-A
theory of weak interactions did not provide an {\it a priori} 
reason for a choice
in this regard.} 
why ${\rm Q_{electron}=-Q_{proton}}$.

We realized that in order to meet these {\it four
aesthetic demands}, the following rather unconventional 
ideas would have to be introduced: 

(i) First, one must place quarks and leptons within the same 
multiplet and
gauge the symmetry group of this multiplet to generate simultaneously
weak, electromagnetic and strong interactions \cite{JPAS,JCPAS}.

(ii) Second, the most attractive manner of placing
quarks and leptons in the same multiplet, it appeared to us  
\cite{JPAS}, was to assume that quarks do possess the 
SU(3)-color degree of freedom, and 
to
extend SU(3)-color to the symmetry SU(4)-
color, interpreting lepton number as the fourth color.  
{\it A dynamical unification of quarks
and leptons is thus provided by gauging the full symmetry SU(4)-color.}  
The spontaneous
breaking of SU(4)-color to $SU(3)^c \times U(1)_{B-L}$ at a sufficiently
high mass-scale, which makes leptoquark gauge bosons superheavy, was then
 suggested to explain the apparent distinction between quarks
and leptons, as regards their response to strong interactions at low
energies.  Such a distinction should then disappear at appropiately high
energies.  

Within this picture, one had no choice but to view {\it fundamental}
strong interactions of quarks as having their origin entirely
in the octet of gluons associated with the SU(3)-color gauge
symmetry  In short,
as a {\it by-product} of our attempts to achieve a higher unification 
through SU(4)-color,
we were led to conclude
that low energy electroweak and fundamental
strong interactions must be generated by the {\it combined gauge
symmetry} $SU(2)_L \times U(1)_Y \times SU(3)^C$, which now
constitutes the symmetry of the standard model \cite{JPAS,HNSU3,FGM}.  
It of course contains the electroweak symmetry
$SU(2)_L \times U(1)_Y$ \cite{SU(2)}.  The idea of the SU(3)-color gauge 
force
became even more compelling with the discovery of asymptotic freedom
about nine months later \cite{GrWzk}, which explained approximate scaling
in deep inelastic ep-scattering, observed at SLAC.

(iii) Third, it became clear that 
together with SU(4)-color one must gauge the commuting
left-right symmetric gauge structure $SU(2)_L \times SU(2)_R$, rather
than $SU(2)_L \times U(1)_{I_{3R}}$, so that electric charge is
quantized.  In short the route to higher unification should include 
{\it minimally} the
gauge symmetry \cite{JPAS,JCPAS}
\begin{equation}
{\rm 
G(224) = SU(2)_L \times SU(2)_R \times SU(4)^C}
\end{equation}
with respect to which all members of the electron-family 
fall into the neat pattern:
\begin{equation}
{\rm
F^e_{L,R} = \left[
\begin{array}{llll}
u_r & u_y & u_b & \nu_e \\
d_r & d_y & d_b & e^-
\end{array}
\right]_{L,R}}.
\end{equation}

With respect to G(224), the left-right-conjugate multiplets
$F_L^e$ and $F_R^e$ transform as (2,1,4) and (1,2,4) respectively; likewise for the mu and
the tau families.

Viewed against the background of particle physics of 1972, as mentioned
above
the symmetry structure G(224) brought
some attractive features to particle physics for the first time.  
They are:

(i)  Organization of all members of a family ($8_L + 8_R$) within
one left-right self-conjugate multiplet, with their peculiar
hypercharges fully explained.

(ii)  Quantization of electric charge, explaining why ${\rm Q_{electron}
= -Q_{proton}}$.

(iii)  Quark-lepton unification through SU(4)-color.

(iv)  \underline{Left-Right and Particle-Antiparticle Symmetries 
in the Fundamental Laws}: With the left-right symmetric gauge structure 
${\rm SU(2)_L \times
SU(2)_R}$, as opposed to ${\rm SU(2)_L \times U(1)_Y}$, 
it was natural to postulate that at the deepest level nature
respects parity and charge conjugation, which are violated only
spontaneously \cite{JCPAS,RMJP}.  Thus, within the symmetry-structure 
G(224), quark-lepton distinction and parity
violation may be viewed as low-energy phenomena which should disappear
at sufficiently high energies.

(v) \underline{Existence of Right-Handed Neutrinos}: Within
G(224), there must exist the right-handed (RH) neutrino ($\nu_R$),
accompanying the left-handed one ($\nu_L$), for each family, because
$\nu_R$ is the fourth color - partner of the corresponding RH up-
quarks.  It is also the $SU(2)_R$ -doublet partner of the associated RH
charged lepton (see eq. (2)).  The RH neutrinos seem to be
essential now (see later discussions) for understanding the non-
vanishing light masses of the neutrinos, as suggested by the recent
observations of neutrino-oscillations.

(vi)  \underline{B-L as a local Gauge Symmetry}:
SU(4)-color introduces B-L as a local gauge symmetry.  Thus, following
the limits from E$\ddot{o}$tvos experiments, one can argue that B-L must be
violated spontaneously.    It has been realized, in the 
light of recent
works on electroweak sphaleron effects, that such spontaneous violation
of B-L may well be needed to implement baryogenesis via leptogenesis 
\cite{Kuzmin}.

(vii) \underline{Proton Decay: The Hall-Mark of Quark-Lepton
Unification}:

We recognized that the spontaneous violation of B-L, mentioned above, is
a reflection of a more general feature of non-conservations of baryon
and lepton numbers in unified gauge theories, including those going
beyond G(224), which group quarks and leptons in the same multiplet
\cite{JCPAS,JCPBL}.  Depending upon the nature of the gauge symmetry and
the multiplet-structure, the violations of B and/or L could be either
spontaneous \footnote{The case of spontaneous violation arises because 
the massless gauge particle coupled to any linear combination of B and L
(which is gauged) must acquire a mass through SSB in order to conform
with the limits from the E\"{o}tvos type experiments.  The corresponding
charge (B and/or L) must then be violated spontaneously.}
, as is the case for 
the non-conservation of B-L in SU(4) color, and those of
B and L 
in the maximal one-family symmetry like SU(16) \cite{SU16};
alternatively, the violations could be explicit, which is what happens
for the subgroups of SU(16), like
SU(5) \cite{SU5} or SO(10) \cite{SO10} (see below). 
One way or another baryon and/or lepton-conservation laws cannot be
absolute, in the context of such higher unification.  The simplest
manifestation of this non-conservation is proton decay ($\triangle B
\neq 0, \triangle L \neq 0$); the other is the Majorana mass of the RH
neutrinos ($\triangle B = 0, \triangle L \neq 0$), as is encountered in
the context of G(224) or SO(10).  An unstable proton thus emerges as the
crucial prediction of quark-lepton unification \cite{JCPAS,SU5}.  Its
decay rate would of
course depend upon more details including 
the scale of such higher unification.

\subsection*{\bf IIB. Going Beyond G(224): SO(10) and SU(5)}

To realize the idea of a single gauge coupling governing the three forces
\cite{JPAS,SU5}, one must embed the standard model symmetry, or G(224),
in a simple or effectively simple group (like SU(N) $\times$ SU(N)).
Several examples of such groups have been proposed.  Howard 
Georgi and Sheldon Glashow
proposed the first such group SU(5) \cite{SU5} which embeds the standard
model symmetry, but not G(224).  Following the discovery of asymptotic
freedom of nonabelian gauge theories \cite{GrWzk} and the suggestion of
SU(5), Georgi, Helen Quinn and Weinberg showed how renormalization effects,
following spontaneous breaking of the unification symmetry, can account
for the observed disparity between the three gauge couplings at low
energies \cite{19}.  Each of these contributions played a crucial role
in strengthening the ideas of higher unification.

To embed G(224) into a simple group, it may be noted that 
it is isomorphic to SO(4) $\times$ SO(6).  Thus the smallest simple
group to which it can be embedded is SO(10) \cite{SO10}.  By the time
SO(10) was proposed, all the advantages of G(224) [(i) to (vi), listed
above] and the ideas of higher unification were in place.  
Since SO(10) contains G(224), the features  
(i) to (vi) are of course retained by SO(10).  In
addition, the 16-fold left-right conjugate set 
(${\rm F_L^e + {\bar{F}}^e_R
}$)
of G(224) corresponds to 
the spinorial 16 of SO(10).  {\it Thus, SO(10) preserves even
the 16-plet family-structure of G(224), without a need for any 
extension.}  If one extends G(224) to the still higher
symmetry ${\rm E_6}$ \cite{E6}, the advantages (i) to (vi) are retained,
as in SO(10), 
but in this case, one must extend the family-structure from a 16 to a
27-plet.

Some distinctions between SU(5) on the one hand 
versus G(224) or SO(10) on the other hand are worth
noting.  Historically, SU(5) served an important purpose, being the
smallest symmetry that embodies the essential ideas of higher
unification.  However, it split members of a family into two
multiplets: $\bar{5}+10$.  By contrast, SO(10) groups all 16 members of
a family into one multiplet.  Likewise, G(224), subject to the
assumption that parity is a good symmetry at high energies, groups the
16 members into one L-R self-conjugate multiplet.  Furthermore, in
contrast to G(224) and SO(10), SU(5) violates parity explicitly from the
start; it does not contain SU(4)-color, and therefore does not possess
B-L as a local symmetry; and the RH neutrino is not an integral feature
of SU(5).  As I discuss below, these distinctions turn out to be
especially relevant to considerations of neutrino masses.

Comparing G(224) with SO(10), as mentioned above, SO(10) possesses
 {\it all} the
features (i) to (vi) of G(224),  but in
addition it offers gauge coupling unification.  I should, however,
mention at this point that the perspective on coupling
unification and proton decay changes considerably in the context of
supersymmetry and superstrings.  In balance, a string-derived G(224)
offers some advantages over a 
string-derived SO(10), while the reverse is true as well.  Thus,
 it seems that a definite choice of one
over the other is hard to make at this point.  I will return to this
discussion in Secs. 4 and 5.

\pagebreak
\section*{\bf	III.	Neutrino Masses: Evidence in Favor of the G(224)
Route to Higher Unification}

Leaving aside the differences between alternative routes to higher
unification, based purely on aesthetic taste, it was of course not clear
in the early 70's as to whether the special features of G(224) --- i.e.
SU(4)-color, left-right symmetry and the RH neutrino --- are utilized by
nature.  
The situation has, however, changed owing to the recent
SuperKamiokande (SK) discovery of the oscillation of ${\rm \nu_\mu}$ to
${\rm \nu_\tau}$ (or $\nu_X$), with a value of $\delta m^2 \approx
\frac{1}{2} (10^{-2} - 10^{-3}) eV^2$ and an oscillation-angle 
$sin^2 2 \theta > 0.82$ 
\cite{SK}.  One
can argue (see e.g. \cite{JPSK}) that the SK result, especially the value
of $\delta m^2$, clearly points to the need for the existence of the RH
neutrinos, accompanying the observed LH ones.  If one then asks the
question: What symmetry on the one hand dictates the existence of the RH
neutrinos, and on the other hand also ensures quantization of electric
charge, together with quark-lepton unification, one is 
led to two very beautiful conclusions: (i) quarks and
leptons must be unified minimally within the symmetry SU(4)-color, and
that, (ii) deep down, the fundamental theory should possess a left-right
symmetric gauge structure: ${\rm SU(2)_L \times SU(2)_R}$.  In short,
the standard model symmetry must be extended minimally to G(224).

One can now obtain an estimate for the mass of $\nu_L^\tau$ in the
context of G(224) or SO(10) by using the following three steps
\cite{JPSK}:

(i) First, assume that B-L and $I_{3R}$, contained in a string-derived
G(224) or SO(10), break near the unification-scale: 
\begin{equation}
{\rm M_X \sim 2 \times
10^{16} GeV},
\end{equation}
 through Higgs multiplets of the type suggested by string-solutions
\cite{String} --- i.e. 		$< (1,2,4)_H >$ 
for G(224) or $< \bar{16}_H >$ for
SO(10).  (The "empirical" determinations of $M_X$ and the new
perspective on unification due to supersymmetry as well as superstrings
are discussed in the next section).  In the process, the RH neutrinos
($\nu_R^i$), which are singlets of the standard model, can and
generically will acquire superheavy Majorana masses of the type
$M_R^{ij} \nu_R^{iT} C^{-1} \nu_R^{j^.}$, by utilizing the VEV of
$< {\bar{16}}_H >$ and effective
couplings of the form:
\begin{equation}
\rm {\cal L}_M (SO(10)) = \lambda^{ij}_R 16_i \cdot 16_j \overline{16}_H
\cdot \overline{16}_H/M_{P \ell} + hc
\end{equation}

A similar expression holds for G(224).  Here i,j=1,2,3, correspond
respectively to $e,\mu$ and $\tau$ families; $M_{pl}$ denotes the
reduced Planck mass $\simeq 2 \times 10^{18} GeV$.  Such gauge-invariant 
non-renormalizable couplings might be expected to be induced by Planck-scale
physics.  (They may well arise - in part or dominantly - by
renormalizable interactions through tree-level exchange of superheavy
states, such as those in the string tower).  Assuming that the Majorana
couplings are family-hierarchical, $\lambda_{33}$ being the leading one,
somewhat analogous to those that give the Dirac masses, and ignoring the
effects of off-diagonal mixings (for simplicity), one obtains:
\begin{equation}
M_{3R} \approx \frac{\lambda_{33} <\bar{16}_H>^2 }{2 \times
10^{18} GeV}
\approx \lambda_{33} (4.5 \times 10^{14} GeV) \eta^2
\end{equation}

This is the Majorana mass of the RH tau neturino.  Guided by
the value of $M_X$, in this 
estimate, we have substituted $< \bar{16}_H > =
(3 \times 10^{16} GeV) \eta$ where $\eta \approx 1/2$ to 2. 

(ii) Second, assume that the effective gauge symmetry below the string-
scale contains SU(4)-color.  Now using SU(4)-color and the Higgs
multiplet $(2,2,1)_H$ of G(224) or equivalently $10_H$ of SO(10), one
obtains the relation $m_\tau(M_X) = m_b(M_X)$, which is known to be
successful.
Thus, there is a good reason to believe that the third family gets 
its masses primarily from the $10_H$ or equivalently $(2,2,1)_H$.  In
turn, this implies:
\begin{equation}
\rm m(\nu^\tau_D) \approx m_{top}(M_X) \approx (100-120) GeV
\end{equation}

(iii) Given the superheavy Majorana masses of the RH neutrinos as well
as the Dirac masses, as above, the see-saw mechanism \cite{SeeSaw}
yields naturally light masses for the LH neutrinos.  For $\nu_L^\tau$
(ignoring mixing), one thus obtains, using eqs. (5) and (6),

\begin{equation}
\rm m(\nu^\tau_L) \; \approx \; \frac{m(\nu^\tau_{Dirac)^2}}{M_{3R}}
\approx (1/45) eV (1 \; to \; 1.44)/\lambda_{33} \eta^2
\end{equation}

Considering that on the basis of the see-saw mechanism, we naturally
expect that $m(\nu^e_L) \ll m(\nu^\mu_L) \ll 
m(\nu^\tau_L)$, and assuming that the SuperK
observation represents $\nu_L^\mu - \nu_L^\tau$ (rather than $\nu_L^\mu
-\nu_X$) oscillation, so that the observed $\delta m^2 \approx 1/2(10^{-
2}-10^{-3})eV^2$ corresponds to $m(\nu_L^\tau)_{obs} \approx$ (1/15 to
1/40) eV, it seems {\it truly remarkable} that the expected magnitude
of $m(\nu_L^\tau)$, given by eq.(7), is just about what is observed, if
$\lambda_{33} \eta^2 \approx$ 1 to 1/3.  Such a range of
$\lambda_{33}\eta^2$ seems most plausible and natural (see discussion in
Ref. [22]).  It should be stressed that the estimate (7) utilizes the
ideas of supersymmetric unification, especially in getting the scale of
$M_X$ (eq.(3)), and of SU(4)-color in getting $m(\nu^\tau_{Dirac})$
(eq.(6)).  The agreement between the expected and the SuperK result thus
suggests that, {\it at a deeper level, near the string or the coupling
unification scale $M_X$, the symmetry group G(224) and thus the ideas of
SU(4)-color and left-right symmetry are likely to be relevant to
nature.}

By providing clear support for G(224), the SK result 
selects
out SO(10) or $E_6$ as the underlying grand unification symmetry, rather
than SU(5).  Either SO(10) or $\rm E_6$ or
both of these symmetries ought to be relevant at
some scale, and in the string context, as discussed later, 
that may well be in
higher dimensions, above the compactification-scale, below which there
need be no more than just the G(224)-symmetry.  If, on the other hand,
SU(5) were regarded as a fundamental symmetry, first, there would be no
compelling reason, based on symmetry alone, to introduce a $\rm \nu_R$,
because it is a singlet of SU(5).  Second, even if one did introduce
$\rm \nu^i_R$ by hand, the Dirac masses, arising from the coupling $\rm
h^i \overline{5}_i < 5_H > \nu^i_R$, would be unrelated to the up-flavor
masses and thus rather arbitrary (contrast with eq. (6)).  So also
would be the Majorana masses of the $\nu^i_R$'s, which are
SU(5)-invariant and thus can even be of order Planck scale (contrast
with Eq. (5)).  This would give $\rm m(\nu^\tau_L)$ in gross conflict
with the observed value.  In this sense, the SK result appears to disfavor
SU(5) as a fundamental symmetry, with or without supersymmetry.

Finally, it is intriguing to note that the SuperK result agrees well with
the idea of supersymmetric unification.  For this purpose, one
could use the mass of $m(\nu_L^\tau)$, suggested by the SuperK data, as
an input to obtain the VEV of $< \bar{16}_H >$, that breaks B-L, as an
output.  By reversing the steps in going from eq. (7) together with eqs.
(6) and (5), one obtains, as is to be expected, $< \bar{16}_H > \sim 3
\times 10^{16}$ GeV (if $\lambda_{33} \sim O(1)$).  It is rather striking 
that
this
is just about the same as the scale of the meeting of the three gauge
couplings, which is obtained from extrapolation of their measured value
at LEP, in the context of supersymmetry (see next section).  {\it In
short, two very different considerations --- light neutrino masses on
the one hand, and gauge coupling meeting on the other hand --- point 
to
one and the same scale for the underlying new physics!}
If one assumes supersymmetric unification, one can hardly avoid noticing
how
beautifully it makes the picture hang together!

In the foregoing, I have discussed only the mass of $\nu_\tau$ in the 
context of G(224) or SO(10). 
In the last section, I will mention briefly how, by adopting familiar
ideas of understanding cabibbo-like mixing angles in the quark-sector,
one can quite plausibly obtain not only the right magnitude for the mass
of $\nu_\tau$ but also a large $\nu_L^\mu - \nu_L^\tau$ oscillation
angle, as observed at SuperK.  I will also discuss that 
simultaneously one can attribute the solar
neutrino-deficit to $\nu_e-\nu_\mu$ oscillation.

I now present the issues associated with coupling unification. 

\section*{\bf IV. Coupling Unification: A New Perspective 
Due To Supersymmetry and Superstrings}

It has been recognized from the early 70's, that the concept of higher 
unification --- now commonly called grand unification --- has two
dramatic consequences:
(i) meeting of the gauge couplings at a high scale, and (ii) proton
decay \cite{JPAS, JCPAS, SU16, 19}.  Equally dramatic is the prediction
of the light neutrino masses, which is a
special feature of only a subclass of grand unification symmetries
that contain SU(4)-color, like SO(10) or $E_6$.  As discussed above, 
this feature 
seems to be borne out by the SuperKamiokande result on neutrino-
oscillations.  The status of the first two predictions are discussed
in this section and the next.  

{\bf\subsection*{IVA. Meeting of The Three Gauge Couplings and  
The Need for Supersymmetry}}

It has been known for some time that the precision measurements of the
standard model coupling constants (in particular ${\rm sin^2 \theta_W)}$ 
at
LEP put severe constraints on the idea of grand unification.
Owing to these constraints, the non-supersymmetric minimal $SU(5)$, and
for similar reasons, the one-step breaking minimal non-supersymmetric
$SO(10)$-model as well, are now excluded.\cite{LGR}  For example, minimal
non-SUSY $SU(5)$ predicts:  $sin^2 \theta_W(m_Z)) \mid_{\bar{MS}} = .214 \pm
.004$, where as current experimental data show:  $sin^2
\theta_W{(m_Z)_{expt}}^{LEP} = .2313 \pm .0003$.  The
disagreement with respect to $sin^2 \theta_W$ is reflected most clearly
by the fact that the three gauge couplings $(g_1,g_2$ and $g_3)$,
extrapolated from below, fail to meet by a fairly wide margin in the
context of minimal \underline{non-supersymmetric} $SU(5)$ (see fig. 1).

But the situation changes radically if one assumes that the standard
model is replaced by the minimal supersymmetric standard model (MSSM),
above a threshold of about $1 TeV$.  In this case, the three gauge
couplings are found to meet\cite{LLUO}, at least
approximately, provided $\alpha_3(m_Z)$ is not too low (see
figs. 2a and 2b).  Their scale of meeting is given
by
\begin{equation}
M_X \, \approx \, 2 \, \times \, 10^{16} GeV \quad ({\rm MSSM \; or \; 
SUSY  SU(5)})
\end{equation}
$M_X$ may be interpreted as the scale where a supersymmetric grand unification 
symmetry (GUT) (like
minimal SUSY SU(5) or SO(10)) --- breaks spontaneously into the
supersymmetric standard model symmetry $SU(2)_L \, \times \, U(1)
\, \times \, SU(3)^c$.

{\it The dramatic meeting of the three gauge couplings
(Fig. 2) thus provides a strong support for both grand
unification and supersymmetry.}

Considering (a) that a straightforward meeting of the three gauge
couplings occurs, only provided supersymmetry is assumed; (b) that
         supersymmetry provides at least a technical resolution
         of the gauge hierarchy problem, by preserving the small 
input value of the ratio
of ($m_W/M_X$), in spite of quantum corrections; and
         (c) that it is needed for consistency of string theory, it
         seems apparent that {\it supersymmetry is an essential
         ingredient for higher unification}.

{\bf \subsection*{IVB.	The Issue of Compatibility Between MSSM and String
Unifications}}

The idea of grand unification would be incomplete without incorporating
the unity of gravity with the weak, electromagnetic and the strong QCD
forces.  Superstring theory \cite{Superstring}, and now the M theory
\cite{MTheory} provide however the only known framework that exhibits
the scope for such a unity.  It thus becomes imperative that the meeting
of the gauge couplings of the three non-gravitational forces, which
occur by the extrapolation of the LEP data in the context of MSSM, be
compatible with string unification.

Now, string theory does provide gauge coupling unification for the
effective gauge symmetry, below the compactification-scale.  The new
feature is that even if the effective symmety is not simple, like SU(5)
or SO(10), but instead is of the form G(213) or G(224)
 (say), the gauge couplings of G(213) or G(224) should still
exhibit familiar 
unification at the string-scale, for compactification involving
appropriate Kac-Moody levels (i.e. $k_2=k_3=1$, $k_Y=\frac{5}{3}$ 
for G(213)), barring of course string-threshold corrections
\cite{Ginsparg}.  And even more, the gauge couplings unify with the
gravitational coupling ($\frac{8 \pi G_N}{\alpha'}$) as well at the 
string scale, where 
$G_N$ is
the Newton's constant and $\alpha'$ is the Regge slope.

Thus one can realize coupling unification without having a GUT-like
symmetry below the compactification scale.  This is the {\it new
perspective} brought forth by string theory.  There is, however, an
issue to be resolved.  Whereas the MSSM-unification scale, obtained by 
extrapolation of low energy data is given by $M_X \approx 2 \times
10^{16}$ GeV, the expected one-loop level string-unification scale
\cite{Ginsparg} of $M_{st} \approx g_{st} \times (5.2 \times 10^{17}
GeV) \approx 3.6 \times 10^{17}$ GeV is about twenty times higher 
\cite{WRev,Diennes}.  Here, one has used $\alpha_{st} \approx 
\alpha_{GUT}$ (MSSM) $\approx$ 0.04.  

A few alternative suggestions which have been proposed to remove this
mismatch by nearly a factor of 20 between $M_X$ and $M_{st}$, are as
follows:

{\bf Matching Through String-Duality:}
One suggestion in this regard is due to
Witten \cite{Witten}.  Using the equivalence of the
strongly coupled heterotic $SO(32)$ and the $E_8 \times E_8$ superstring 
theories in 
$D=10$, respectively to the weakly coupled $D=10$ Type I and an
eleven-dimensional $M$--theory, 
he observed that the
4-dimensional gauge coupling and $M_{\rm st}$ can both be
small, as suggested by MSSM extrapolation of the low energy data, without
making the Newton's constant unacceptably large.  

{\bf Matching Through String GUT:}  A second way in which the
mismatch between $M_X$ and $M_{\rm st}$
could be resolved is if superstrings yield an intact supersymmetric
grand unification
symmetry like $SU(5)$ or $SO(10)$ with the right spectrum -- i.e.,
three chiral families and a suitable Higgs system $M_{\rm st}$ \cite{StGUT}
, and if this symmetry
would break spontaneously at $M_X \approx (1/20~ {\rm to}~ 1/50)
M_{\rm st}$ to the standard model symmetry.  However, as yet, there is
no realistic, or even close-to realistic, string--derived GUT model 
\cite{StGUT}.
In particular, to date, no string-derived 
solution exists with a resolution of the 
doublet-triplet splitting problem,  without which one faces 
the problem of rapid proton decay (see discussions later).

{\bf Matching Through Intermediate Scale Matter:}  A third
alternative is based on string--derived standard
model--like gauge groups.  It attributes the mismatch between $M_X$ and
$M_{\rm st}$ to the existence of new matter with intermediate
scale masses ($\sim 10^9-10^{13}$ GeV), which may emerge from
strings \cite{SMLike}.  Such a resolution is
in principle possible, but it would rely on the delicate balance
between the shifts in the three couplings and on the existence of very
heavy new matter which in practice cannot be directly tested by experiments.

{\bf Matching Through ESSM -- A Case for Semi-Perturbative Unification
:}

Babu and I suggested that a resolution of the mismatch between $M_X$ and 
$M_{st}$ 
can come about if there exists two "light" vector-like families (16 + 
$\bar{16}$) 
at the TeV scale \cite{Babu}.  Such a spectrum has an apriori motivation 
in that it provides a simple reason 
for inter-family mass-hierarchy. It can also be tested at LHC.  
Including two \cite{Babu} 
and even three-loop 
effects \cite{JMR}, 
this spectrum leads to a semi-perturbative
unification, with $\alpha_{GUT} \approx .2 - .3$, and 
raises $M_X$ to (1 - 2) $\times 10^{17}$ GeV.  Such
higher values of $\alpha_{GUT}$ (compared to .04 for MSSM) may provide an 
additional advantage by helping to stabilize the dilaton.  

While each of the solutions mentioned above possesses a certain degree
of plausibility (see Ref. 31 for some additional possibilities), 
it is far from clear which, if any, is utilized by the
true string-vacuum.  This is of course related to the fact that, as yet,
there is no insight as to how the vacuum is selected in the string or
in the M-theory.

In summary, string theory, as well as M-theory, fully retain the basic
concept of grand unification --- i.e. unification of matter and of its
gauge forces.  But they enrich the scope considerably by (a) unifying
all matter of spins 0, 1/2, 1, 3/2, 2 and higher, and (b) unifying
gravity with the other forces.  As noted above, the perspective on gauge
coupling unification however changes in the string context, because such
a unification can occur at the string scale, even without having a GUT-
like symmetry at that scale.  In the next section, I discuss the
advantages as well as possible disadvantages of GUT versus non-GUT
string solutions, keeping in mind the issues of both coupling
unification and rapid proton decay.
 
I now turn to considerations of proton decay.

\section*{\bf {V.	Proton Decay as a Probe to Higher Unification}}

{\large\bf VA.} As mentioned before, one of the hallmarks of 
grand unification is
         non-conservation of baryon and lepton numbers, which for most
         simple models, lead to proton decay \cite{JCPAS,SU5}.  
The general
         complexion of baryon and lepton number non-conserving
         processes, including alternative modes of proton decay,
         $n-\bar{n}$ oscillation and neutrinoless double beta decay is
         discussed in my talk at the Oak Ridge Conference  
\cite{JPOak}.  Here I will
         focus on proton decay.

Almost 25 years have passed since the suggestion of proton decay 
was first made in the context of unified theories, in 
         1973.  While there was considerable resistance from the
         theoretical community against such ideas at that point, the 
psychological barrier against them softened over the years.  
The growing interest in the prospect of such a
         decay thus led to the building of proton-decay detectors
         in different parts of the world, 
including the most sensitive one of the
         80's (IMB) at Cleveland, followed by Kamiokande in Japan.
         While proton decay is yet to be discovered, it is encouraging that
         searches for this decay continues at SuperKamiokande with
         higher sensitivity than ever before and detectors such as
         ICARUS are planned to come.  The dedicated searches for proton
         decay at IMB (which was operative till a few years ago) and
         Kamiokande \cite{SOak} already put severe constraints on grand
         unification for over a decade.  Owing to these constraints, the
         non-supersymmetric minimal SU(5) and the minimal SO(10) models
         as well (with one-step breaking) are now excluded.
         In particular, conservatively, minimal non-SUSY SU(5)
         predicts: $\Gamma(p \rightarrow e^+ \pi^0)^{-1} \leq
         (6-10) \times 10^{31}$ yr, where as current data including those
         from Superkamiokande \cite{SKP} yields:
\begin{equation}
\Gamma(p \rightarrow e^+ \pi^0)^{-1}_{expt}>1.6 \times 10^{33} yr.
\end{equation}

\subsection*{{\bf  VB.  The Issue of Proton-Longevity in SUSY Grand 
Unification}}

                Although non-supersymmetric minimal SU(5) or SO(10) are
         excluded by proton-decay searches, as well as by precision
         measurements of ${\rm sin^2 \theta_W}$, 
the situation with regard to
         both issues alters radically, once supersymmetry is combined
         with the idea of grand unification.  First, as mentioned
         before, SUSY makes it possible for the three gauge couplings to
         meet at a common scale M$_X \approx 2 \times 10^{16}$ GeV.  
If one uses $\alpha_3$ and $\alpha_2$ as
         inputs, it correspondingly leads to the correct prediction for 
${\rm sin^2 \theta_W}$.

As regards proton decay, supersymmetric grand unified
         theories (GUTS), bring two new features: (i)   First, by
         raising M$_X$ to a higher value compared to the
         non-supersymmetric case, as above, they strongly suppress the
         gauge-boson -mediated d=6 proton decay operators, so that one
         obtains $\Gamma(p \longrightarrow e^+ \pi^0)^{-1}_{d=6} \approx
         10^{36 \pm 1.5}$ yr.  This is of course compatible with current
         experimental 
limits (eq(9)).  (ii)  Second, they generate d=5 proton decay
         operators of the form $Q_i Q_j Q_k L_\ell$/M  and $\Ubar \Ubar
\Dbar \Ebar$
in the
         superpotential, through the exchange of color triplet
         Higgsinos, which are the GUT partners of the electroweak
         Higgsino doublets \cite{WS}.  These triplets lie, for example,
         in the 5($\bar{5}$) of SU(5), or in the 10 or SO(10).  Since the
         corresponding amplitudes are damped by just one power of the
         mass of the color-triplet higgsinos(m$_{H_c}$), these d=5
         operators provide the dominant mechanism for proton decay in
         supersymmetric GUT.

The d=5 operators have marked effects both on the branching ratios of
different decay modes as well as on the rate of proton decay.  First,
owing to (a) color-antisymmetry, (b) Bose symmetry of the
scalar squark and slepton fields, and (c) the family-hierarchical Yukawa
couplings, it turns out that these d=5 operators (to be called
"standard" d=5) lead to dominant antineutrino modes:
\begin{equation}
p \rightarrow \bar{\nu}_\mu K^+, \bar{\nu}_\mu\pi^+ (standard \hspace{2mm}
  d=5),
\end{equation}
but highly suppressed $e^+\pi^0$, $e^+K^0$ and even $\mu^+\pi^0$ and
$\mu^+K^0$ modes (at least for small and moderate $\tan \beta \leq 15$).
Recall, by contrast, that for non-supersymmetric GUTS, $e^+\pi^0$ is
expected to be the dominant mode.  

Second, given the Yukawa couplings of the electroweak Higgs doublets
(inferred from fermion masses), a typical contribution to the standard
d=5 proton decay operator of the form QQQL/M is found to have an
effective strength $\approx$ ($m_c m_s \sin \theta_c/ v_u v_d$)
($1/M_{H_c}$) $\approx \frac{10^{-7} tan \beta}{M_{H_c}}$ at the GUT-scale.
  Now, for
plausible values or limits 
on $m_{\tilde{q}} \leq 1$ TeV, ($m_{\tilde{W}}/m_{\tilde{q}})
\geq 1/6$ and $\tan\beta \geq 3$ (say), the d=5 operator, as noted above,
subject to wino-dressing,
leads to an inverse decay rate \cite{Hisano}
\begin{equation}
\Gamma^{-1} (p \rightarrow {\bar{\nu}_{\mu}} K^+) \leq 3 \times 10^{32}
yrs (\frac{M_{H_c}}{3 \times 10^{16} GeV})^2
\end{equation}
To be conservative, this estimate uses the minimum theoretical value
of the hadronic matrix element ($\beta_H = .003 GeV^3$), and assumes a
cancellation by a factor of two betwen $\tilde{t}$ and $\tilde{c}$ -
contributions, (although, in general, one could gain a factor of 2 to
4 (say) in the rate on each count).  Given the current experimental
limit of $\Gamma(p \rightarrow \bar{\nu}K^+)^{-1} > 5.5 \times 10^{32}$
yrs ($90 \%$ CL) \cite{Takita}, 
it follows that the color-triplets must be superheavy.
Conservatively \cite{UUDE},
\begin{equation}
M_{H_c} \geq (3-5) \times 10^{16} GeV
\end{equation}
While the color triplets need to be superheavy, their doublet-partners
must still be light ({$\leq$ 1 TeV).  The question arises: 
How can the color-
triplets become superheavy, while the doublet-partners remain naturally
light?  This is the well-known problem of {\it doublet-triplet
splitting} that faces all SUSY GUTS.

  Leaving out the possibility of extreme fine
tuning, two of the proposed solutions to this problem are as follows:

(i) \underline{The Missing Partner Mechanism} \cite{Missing}:  In
this case, by introducing suitable large-size Higgs multiplets, such as
$50_H \, + \, \overline{50}_H \, + \, 75_H$, in addition to $5_H \, + \,
\overline{5}_H$ of $SU(5)$, and introducing couplings of the form $W \, =
\, C \, 5_H \, \cdot \, \overline{50}_H \, \cdot \, <75_H> \, + D \,
\overline{5}_H \, \cdot \, 50_H <75_H>$, one can give superheavy masses
to the triplets (anti-triplets) in $5(\overline{5})$ by pairing them with
anti-triplets (triplets) in $\overline{50}$(50).  But there do not exist
doublets in $50(\overline{50})$ to pair up with the doublets in
$5(\overline{5})$, which therefore remain light.
\newline (ii) \underline{The Dimopoulos-Wilczek Mechanism} \cite{DW}:
Utilizing the fact that the VEV of $45_H$ of $SO(10)$ does not have to
be traceless (unlike that of $24_H$ of $SU(5))$, one can give mass to
color-triplets and not to doublets in the $10$ of $SO(10)$, by arranging
the $VEV$ of $45_H$ to be proportional to $i\tau_2 \times$ 
diag $(x, x, x, o, o)$, and
introducing a coupling of the form $\lambda 10_{H1} \, \cdot \, 45_H \,
\cdot \, 10_{H2}$ in $W$.  Two $10's$ are needed owing to the
anti-symmetry of $45$.  Because of two $10's$, this coupling would leave
two pairs of electroweak doublets massless.  One must, however, make one
of these pairs superheavy, by introducing a term like $M_{10}10_{H2}
\cdot 10_{H2}$ in
$W$, so as not to spoil the successful prediction of $\sin^2\theta_W$ of
SUSY GUT.  In addition, one must
also ensure that only $10_{H1}$ but not $10_{H2}$ couple to the light
quarks and leptons, so as to prevent rapid proton decay.  All of these
can be achieved by imposing suitable
discrete symmetries.  There is, however, still some question as to
whether the mass-scale $M_{eff}  
 \equiv 
 (\lambda < 45_H >)^2/M_{10} $ that controls the d=5 amplitude
can be of order $10^{18}$ GeV (that is needed), 
without conflicting with unification of the gauge couplings.

In summary, solutions to the problem of doublet-triplet splitting needing
a suitable {\it choice} of 
Higgs multiplets and discrete symmetries
are technically feasible.  It is however not clear whether any of these
mechansims can be consistently derived from an underlying theory, such
as the superstring theory.
To date, no such mechanism has been realized in
a {\it string-derived} GUT solution \cite{StGUT}.

\subsection*{\bf VC. Rapid Proton Decay And The Other 
Problems of Naturalness in 
Supersymmetry}

In addition to the problem of doublet-triplet splitting that faces 
 SUSY GUT theories, it is important to
note that there is 
a generic problem for all supersymmetric theories, involving either
a GUT or a non-GUT symmetry, in the presence of quantum gravity.
This is because, in accord with the standard model gauge symmetry
$SU(2)_L \times U(1)_Y \times SU(3)^C$, a supersymmetric theory in general
permits, in contrast to non-supersymmetric ones, dimension $4$ and dimension
$5$ operators which violate baryon and lepton numbers \cite{WS}.  Such 
operators are likely to be induced by Planck-scale physics including
especially quantum gravity, unless they are forbidden by symmetries of
the theory.
Using standard notations, the operators in question are as follows:
\bea
W &=& [\eta_1 \Ubar \, \Dbar \, \Dbar  + \eta_2 Q L \Dbar + \eta_3 L L
\Ebar ] \nn\\
&+& [\l _1 QQQL + \l _2 \Ubar \, \Ubar \, \Dbar \, \Ebar  +
\l _3 LLH_2 H_2]/M.
\tea
Here, generation, $SU(2)_L$ and $SU(3)^C$ indices are suppressed.
$M$ denotes a characteristic mass scale.
The first two terms of $d=4$, jointly, as well as the $d=5$ terms of
strengths
$\lambda_1$ and $\lambda_2$, individually,
induce $\Delta(B-L) = 0$ proton decay with amplitudes
$\sim \h _1 \h _2/m_{\tilde
{q}}^{2}$ and $(\l _{1,2}/M)(\d )$ respectively, where $\d $
represents a loop-factor. Experimental limits on proton
lifetime turn out to impose the constraints:
$\h _1 \h _2 \leq 10^{-24}$ and $(\l _{1,2} /M) \leq 10^{-23}$ to $10^{-24} 
GeV^{-1}$.
Thus, even if $M \sim M_{string} \sim 10^{18}$ GeV, we must
have  $\l _{1,2} \leq 10^{-5}$ to $10^{-6}$ ,
so that proton lifetime will be in accord with experimental limits.

Renormalizable, supersymmetric standard-like and $SU(5)$ \cite{Dimop}
models can be constructed so as to avoid,
{\it by choice}, the $d=4$ operators (i.e. the $\h _{1,2,3}$-terms) by
imposing a discrete or a multiplicative
$R$-parity symmetry: $R\equiv (-1)^{3(B-L)}$, or more
naturally, by gauging $B-L$, as in ${\cal G}_{224}
\equiv SU(2)_L\times SU(2)_R\times SU(4)^C$ or $SO(10)$.
Such resolutions,
however, do not in general suffice if we
permit higher dimensional operators and
intermediate or GUT-scale VEVs of fields which violate $(B-L)$
by one unit and thereby $R$-parity (see below). 
In string solutions, VEV's of such fields seem to be needed, to generate
Majorana masses for the RH neutrinos.  Besides, $B-L$ can not
provide any protection against the $d=5$ operators given
by the $\l_1$ and $\l_2$ - terms, which conserve $B-L$. As mentioned 
above these
operators are, however, expected to be present in any
theory linked with gravity, e.g. a superstring theory,
unless they are forbidden by some new symmetries.

These considerations show that, in the
context of supersymmetry, the extraordinary stability of the
proton is a major puzzle.  And, the problem is heightened especially
in the context of SUSY GUT theories because of the need for the 
doublet-triplet splitting in such theories.  
{\it The question in fact arises:  Why does
the proton have a lifetime exceeding $10^{40}$ sec, rather than the
apparently natural value, for supersymmetry, of less than 1 sec?}  As
such, the known longevity of the proton deserves a natural
explanation.  I believe that it is in fact a major clue to 
some deeper physics
that operates near the Planck-scale.  

Apart from the problem of rapid proton decay, supersymmetry in fact 
generates a few additional problems of similar magnitude.  
  These together
constitute the so-called
{\it naturalness
problems of supersymmetry}.  They include understanding: 
(i) the extreme smallness of the SUSY-breaking mass-
splittings compared to the Planck-scale 
(i.e. why ($\delta m_s/M_{planck}$) $\sim 10^{-15}$ rather than 
order unity), (ii) the smallness of the
$\mu$-parameter of MSSM also compared to the Planck-scale,
(iii) the strong suppression of the neutrino-Higgsino
mixing mass, (that needs to be less than about 1 MeV) in a context
where R-parity is violated, and (iv) the smallness of especially the CP-
violating part of the $K^o-\bar{K}^o$ amplitude in spite of the
potentially large contributions from squark and gluino loops.  
In addition to this
set of problems, which are special to supersymmetry, there is of course
the familiar challenge of understanding the hierarchical masses and
mixings of quarks and leptons.  Resolving these problems would amount to 
understanding the origins of some
extremely small numbers, ranging 
from $10^{-6}$ to $10^{-19}$, which apriori
could be of order unity.  As such, 
I believe that they are a reflection
of {\it new symmetries} which operate near the Planck-scale. 
In the limit of 
these symmetries, the respective entities, such as the strengths of the 
d=4 and d=5 operators and the magnitudes of $\delta m_s$ and $\mu$, 
would  vanish.  Although the symmetries break, quite possibly near the 
GUT-scale, they need to be
powerful enough to provide the needed protection up
to sufficiently high order in non-renormalizable terms, scaled by the 
Planck mass, so as to 
render the respective numbers as small as they are.
Symmetries of this nature simply do not exist in
conventional GUTS.  They do, however, arise, not so infrequently, in
string-solutions, including some which are fairly realistic, possessing
three-families and hierarchical Yukawa couplings
\cite{Faraggi,Anto,Flipped}.

Invariably, these solutions possess non-GUT symmetries such as (i) the
(B-L)-preserving standard model-like symmetry G(2113)
\cite{Faraggi}, or (ii) G(224) \cite{Anto}, or (iii) flipped SU(5) $\times$
U(1) \cite{Flipped}.  Based on some recent work \cite{StProton}, I note
below how string symmetries can play an essential role in avoiding the
danger of rapid proton decay and also help in resolving some of the
other naturalness problems noted above.  

\subsection*{\bf VD.  The Role of String-Flavor Symmetries in Resolving
 The Naturalness Problems}

To illustrate the usefulness of string-symmetries, I would consider
especially a class of three-family string solutions which are based on
the free fermionic construction \cite{Fermionic} and correspond to a
special $Z_2 \times Z_2$ orbifold compactification \cite{Faraggi}.  They
lead, after the applications of all GSO-projections, to a gauge symmetry
at the string-scale of the form:
\begin{equation}
{\cal G}_{st} = [SU(2)_L \times SU(3)^C \times U(1)_{I_{3R}}
\times U(1)_{B-L}]
\times [G_M = \prod_{i=1}^{6} U(1)_{i}] \times G_H.
\end{equation}
The first factor will be abbreviated as G(2113).  Here $U(1)_i$ denote
six horizontal symmetries which act non-trivially on the three families
($e,\mu$ and $\tau$) and distinguish between them.  $G_H$ denotes the
hidden-sector symmetry which operate on "hidden" matter.  The horizontal
symmetries $U(1)_i$ couple to both the observable and the hidden sector
matter.  

The crucial point is that the pairs ($U_1, U_4$), ($U_2,U_5$) and ($U_3,
U_6$), respectively couple to families 1, 2 and 3, in an {\it identical
fashion}.  \footnote{While $U_1, U_2$ and $U_3$ respectively assign the
same charge to all 16 members of families 1,2 and 3, $U_4,U_5$ and $U_6$
distinguish between members within a family.  Thus $U_1, U_2$ and $U_3$ commute with SO(10), but $U_4, U_5$
and $U_6$ do not.}  Thus, on the one hand, these six U(1) symmetries,
having their origin in SO(44) \cite{Fermionic}, distinguish between the
three families, unlike a GUT symmetry like SO(10).  Thereby they serve
as generalized {\it "flavor" symmetries} and in turn help explain the
hierarchical Yukawa couplings of the three families \cite{Faraggi}.  On
the other hand, the coupling of the three pairs ($U_1,U_4$), ($U_2,U_5$)
and ($U_3,U_6$) fully preserve the cyclic {\it permutation symmetry}
with respect to the three families.

Turning to the problem of rapid proton decay in the context of these
string solutions, there are two features which together help resolve the
problem.  First, it turns out that for non-GUT solutions of the type
obtained in Ref. [47] (this is also true of the G(224)-solution of Ref.
[48]), in the process of compactification leading to G(2113), the
dangerous color triplets are simply projected out of the spectrum
altogether.  As a result, the problem of doublet-triplet splitting is
neatly avoided.  {\it This is an obvious advantage of a non-GUT over a
GUT string solution.}  

Second, it needs to be said that of the six
U(1)'s [Ref. 47], one linear combination --- i.e. $U(1)_A = 1/\sqrt{15}
[2(U_1 +U_2 + U_3) - (U_4 +U_5 + U_6)]$ --- is anomalous, while the other
five are anomaly-free (occurrence of such anomalous U(1) is in fact
fairly generic in string solutions).  Furthermore, the string solutions
invariably yield a set of standard model singlet fields {
$\{\Phi_a\}$ which
couple to the flavor symmetries $U(1)_i$.  For the solution of Ref. 47,
they couple to the six U(1)'s as well as to B-L and $I_{3R}$.  Now a set
of these $\{\Phi_i\}$ fields must acquire VEV's of order ($10^{-1} - 10^{-
2}$) $M_{pl}$ (where $M_{pl} \approx 2 \times 10^{18}$ GeV), in order to
cancel the Fayet-Iliopoulos D-term generated by $U(1)_A$, and also all F
and D-terms, so that supersymmetry is preserved, barring additional
constraints \cite{Dine}.  

It turns out that the six flavor symmetries
$U(1)_i$, together with certain SUSY-preserving patterns of VEVs of the
$\{\Phi_a\}$-fields, suffice to naturally safeguard proton-longevity, to
the extent needed, from all potential dangers, including those which may
arise through gravity-induced 
higher dimensional operators (d $\geq$4) and the exchange
of color-triplets in the infinite tower of heavy string states
\cite{StProton}.  This protection holds in spite of the fact that
certain $\Phi_i$'s acquiring VEVs carry $\mid{B-L}\mid = 1$, which help
provide superheavy Majorana masses to the RH neutrinos, but, in the
process, break R-parity.  The protection comes about because the 
symmetries mentioned above prevent the appearance of 
the dangerous effective d=4 and d=5 operators, unless one utilizes 
non-renormalizable operators involving sufficiently
high powers of the ratios $< \{\Phi_i\} >/M_{st}$, where each such ratio
is naturally $O(1/10)$.
{\it These virtues of the extra flavor symmetries
show that, believing in supersymmetry, superstring is suggested just to
understand why the proton is so long-lived.}  

In above, I have tried to illustrate the beneficial role of string
symmetries within one class of fairly realistic string solutions
\cite{Faraggi}.  It still remains to be seen whether such 
string-symmetries by themselves can account for the desired suppression
of the d=4 and the d=5 operators, regardless of the choice of the
pattern of VEVs.  [For attempts in this direction, see e.g. Ref.
\cite{SU2} and \cite{Ellis}].  

I should add briefly that the string-flavor symmetries of the type just
described are found to play a crucial role in resolving also some of the
other problems of naturalness listed above.  These include understanding
the smallness of SUSY-breaking mass-splittings ($\delta m_s \sim$ 1 TeV)
on the one hand, and deriving the desired squark-degeneracy that
adequately accounts for the suppression of the flavor-changing neutral
current processes on the other hand.  These two features are realized by
implementing supersymmetry-breaking through a non-vanishing D-term of
the string-derived anomalous U(1) gauge symmetry, noted above
\cite{U1,Dvali}.  The string-flavor symmetries also help in
understanding the strong suppression of the neutrino-higgsino mixing
mass \cite{FJP} and the smallness of the CP violating part of the $K^o -
\bar{K}^o$ amplitude \cite{K}.  Last but not least, the same flavor
symmetries help obtain the qualitatively correct pattern of hierarchical
fermion masses and mixings \cite{Faraggi}.  {\it Thus the beneficial roles of
these string flavor symmetries can hardly be overemphasized. } 

One is of course aware that it is premature to take any specific string
solution, or even a specific class of solutions, from the vast set of
allowed ones, too seriously.  Nevertheless it seems feasible that
certain features, especially the symmetry properties, may well survive
in the final picture that may emerge from the ultimate underlying theory,
encompassing string theory, M theory and D-branes.  These theories may 
of course well generate new symmetries in their strongly interacting phases
which cannot be found in their weakly interacting versions.
 From a purely
utilitarian point of view, given the magnitude
of the naturalness problems, it seems that one way or another
such flavor symmetries {\it should} in fact
emerge from the underlying theory, 
just in order that supersymmetry would not conflict with the ideas
of naturalness.  Here, however, a bottom-up approach seems to be
especially helpful in providing insight into the nature of these flavor
symmetries, that a satisfactory underlying theory needs to produce.

It needs to be mentioned that while the string-flavor symmetries provide
the scope for obtaining a resolution of the problems mentioned above,
obtaining a simultaneous resolution of all or most of them in the
context of a given string solution is still a challenging task. 

\subsection*{\bf VE.	A GUT or a Non-GUT String Solution?} 

In summary, comparing 
string-GUT with non-GUT solutions, where the former yield symmetries
 like SU(5) or SO(10), while the latter lead 
to symmetries like 
G(2113) or
G(224),
at the string scale, we see that each class 
has a certain advantage over the
other.  For a non-GUT solution, the gauge couplings unify only at the
string-scale; thus one must assume that somehow a solution of the type
discussed in Sec. 4 should resolve the mismatch between $M_X$ and $M_{st}$.
This is plausible but not easy to ascertain.  In this regard, a string
GUT-solution yielding SU(5) or SO(10) 
appears to have an advantage over a non-GUT solution,
because, in the case of the former, the couplings naturally stay
together between $M_{st}$ and $M_X$.  Furthermore, a GUT 
symmetry-breaking scale 
of $M_{st}/20$ seems to be plausible in the string-context. 

On the other
hand, as mentioned above, deriving a GUT-solution from strings, while 
achieving doublet-triplet splitting, 
is indeed a major burden, and has not been
achieved as yet.  In this regard,  
the non-GUT solutions seem to possess
a distinct advantage, because the dangerous color-
triplets are often naturally projected out [see e.g. 
\cite{Faraggi},\cite{Anto}].  Furthermore, 
these solutions possess new symmetries, which are not available in GUTS,
and some of these do not even commute with GUT-symmetries, 
but they do  help in providing the desired protection, even against 
gravity-induced proton decay,
that may otherwise be unacceptably rapid \cite{StProton}.  
In addition, as mentioned above, these new symmetries turn out to 
help in the resolution of  
the other naturalness-problems of supersymmetry as well 
(see e.g. Refs.\cite{U1},
\cite{FJP} and \cite{K}).

Weighing the advantages and 
possible disadvantages of both,
it seems hard to make a clear choice between a GUT
versus a non-GUT string-solution.  While one may well 
have a preference
for one over the other, it seems reasonable to keep
one's options open in this regard and look for other means, based e.g.
on certain features of proton decay and the solutions to the naturalness
problems, which can help provide a distinction between the two
alternatives.  Short of making such a choice at this point, one must 
assume
that for a GUT-solution, strings would somehow provide a
resolution of the problem of doublet-triplet splitting, while 
for a non-GUT
string-solution, it needs to be assumed that 
one of the mechanisms mentioned in Sec. 4
(for instance, that based on string-duality \cite{Witten} and/or semi-
perturbative unification \cite{Babu}) is operative so as to
remove the mismatch between
$M_X$ and $M_{st}$.

I now discuss how the masses and the mixings of the fermions, 
especially those of the 
neutrinos, influence proton decay.

\renewcommand{\baselinestretch}{1}
\section*{\bf VI.	Link Between Neutrino Masses and Proton Decay}
\renewcommand{\baselinestretch}{1.4}

Two important characteristics of supersymmetric unification, based on a
gauge symmetry like SO(10) or a string-derived G(224), seem to be borne
out by nature.  They are: (a) gauge coupling unification at a scale $M_X
\sim 2 \times 10^{16}$ GeV, and (b) light neutrino masses ($\ll
m_{e,\mu,\tau}$).  As discussed in Sec. 3, the value of m($\nu_\tau$)
$\approx 1/20$ eV, suggested by the SuperK result, is just about
what one would expect within the symmetry-structure G(224)/SO(10),
if the (B-L)--breaking scale is around $M_X$.  One is
thus naturally tempted to ask: {\it Will proton decay ---the other major
prediction of grand unification --- also reveal in the near future?} 

This question acquires a special significance because of the following
circumstance.  Ordinarily, except for the scale of new physics, involved
in the two cases, proton decay, especially its decay modes are
considered to be essentially unrelated to the pattern of neutrino
masses.  
However, in a recent paper, Babu, Wilczek and I noted that neutrino masses
can have a significant effect on proton decay as regards its rate as well 
as decay modes \cite{BPW1}.  This is because in supersymmetric unified 
theories, based on SO(10) or G(224), assignment of heavy Majorana masses 
to the RH neutrinos (as discussed in Sec. 3), inevitably 
introduces {\it a new set of color-triplets}
(unrelated to the electroweak doublets), 
whose effective couplings to quarks
and leptons are related to these Majorana masses 
(see eqs. (4) and (5)).  Exchange of these new
color-triplets give rise to {\it a new set of d=5 proton decay operators},
which are thus directly related to the neutrino-masses.  These are 
in addition to the {\it standard d=5} operators which arise due to
exchange of the familiar color-triplets that are related to the 
electroweak doublets (see Sec. 5).  Even without 
the SuperK result on atmospheric-oscillation, assuming that
$\rm \nu_e - \nu_\mu$ oscillation is relevant to the MSW explanation of
the solar neutrino puzzle, so that $\rm m(\nu^\mu_L) \approx 3 \times
10^{-3} eV$, which corresponds to $\rm M(\nu^\mu_R) \approx 2 \times 10^{12}$
GeV, the new d=5 operators by themselves (not including contributions
from the standard d=5 operators) lead to proton lifetimes typically in
the range: $\Gamma(p \rightarrow \bar{\nu} K^+)^{-1}_{\rm New Op} \approx 
10^{31.5 \pm 3}$ yrs.   
Now it could happen that the contributions from the standard d=5
operators are somehow suppressed.  In particular, this 
would arise in the
case of non-GUT string solutions leading to symmetries like G(224) or
G(2113) for which the standard color-triplets get projected out through 
compactification \cite{Faraggi, Anto}.  
Even in this case, the new operators related to
neutrino masses can still contribute to proton decay.  As noted above,
these lead to proton lifetimes in an interesting range which is
accessible to SuperKamiokande searches.

Furthermore, 
the flavor-structure of the new d=5 operators are expected to
be distinct from those of the standard d=5 operators, which are governed
by  the highly hierarchical Dirac masses of quarks and leptons.  In
contrast to the standard d=5 operators, the new ones can lead to
prominent charged lepton decay modes, such as
$\ell^+ \pi^o, \ell^+ K^o$ and $\ell^+ \eta$, involving especially 
$\mu^+$,
even for low or moderate values of $tan \beta \leq$ 20, together with 
$\bar{\nu}K^+$-modes.
The intriguing feature thus is that owing to the underlying SO(10)
or just SU(4)-color symmetry, {\it proton decay operator knows about
neutrino masses and vice versa}.

SuperK result introduces new features to proton decay.  
With a {\it maximal} effective Majorana-coupling for the third family (i.e.
$\rm \lambda_{33} \sim {\cal{O}}(1))$, as suggested in Sec.3, 
that corresponds
to $\rm M_{3R} \approx (few \times 10^{14} GeV)$ for the case of no mixing 
(see eq. (5)), 
one might worry that proton would decay too fast, because of
an enhancement in the new d=5 operators, relative to that considered in
Ref. \cite{BPW1}.  It turns out, however, that
because $\rm \tau^+$ is heavier than the proton and also because $\rm
\bar{\nu}_\tau K^+$ mode receives a strong suppression-factor from the
small mixing angle associated with
the third family $(V_{ub} \approx 0.002 - 0.005)$, a
maximal Majorana-coupling of the third family $\rm (\lambda_{33} \sim
{\cal{O}}(1))$, that corresponds to 
$\rm m(\nu^\tau_L) \approx (1/10 - 1/30) eV$, 
leads to dominant (or prominent) $\bar{\nu}_{\tau}K^+$-mode; but such a
coupling is still compatible with present limit on proton lifetime
\cite{BPW2}.

Babu, Wilczek and I have recently attempted
to understand the neutrino masses and mixings, as
suggested by both the SuperKamiokande result 
(interpreted as $\nu_\mu - \nu_\tau$ 
oscillation) and the solar neutrino puzzle, within a predictive {\it
SO(10) or G(224)-based} quark-lepton unified description of the masses
and mixings of {\it all fermions}---i.e. quarks, charged leptons as well as
neutrinos \cite{BPW2}.  Adopting familiar ideas of
generating hierarchical eigenvalues through off-diagonal mixings and
correspondingly Cabibbo-like mixing angles, we find that the bizarre
pattern of masses and mixings of quarks and charged leptons of all three
families can in fact be described adequately (to better than $10 \%$
accuracy), within an economical SO(10)-framework, which makes five
successful predictions,  
just for the quark and the charged lepton system.

In the process, the Dirac mass matrices of the neutrinos, as well as of 
the
charged leptons, get fully determined.  Taking the Dirac
masses, thus fixed, together with a simple
hierarchical pattern for the Majorana mass matrix of the superheavy
right-handed neutrinos, we show that one can obtain quite naturally
a large $\nu_L^\mu -
\nu_L^\tau$ oscillation angle ($sin^2 2\theta \simeq .85-.95$), just as
observed at SuperK, in spite of highly non-degenerate masses of the 
three light
neutrinos---e.g. with m($\nu_L^e$) $\ll$ m($\nu_L^\mu$) $\approx$ 
($\frac{1}{10}-\frac{1}{20}$)m($\nu_L^\tau$), where $m(\nu_L^\tau) \approx$
(1/20 eV)(1/2 - 2).  Such a hierarchical 
mass-pattern for the light neutrinos is of course natural to see-saw.  In this case, $\nu_e -
\nu_\mu$ oscillation becomes relevant to the small angle MSW explanation
\cite{MS} 
of the solar neutrino puzzle \cite{MS2}.  
The distinctive features of this
explanation of the neutrino-anomalies are: 
(a) its origin within an underlying unified theory that
relates the masses and mixings of neutrinos to those of quarks and charged 
leptons, and (b) the emergence
of the large oscillation angle without a large mixing in either the
($\nu_\mu -\nu_\tau$) or the ($\mu-\tau$) mass-eigenstates.

As an important corollary to this work, owing to the link mentioned 
above between
neutrino masses and proton decay \cite{BPW1}, we find that the mass of
 $\nu_\tau$
 and the large oscillation angle
suggested by the SuperKamiokande result in fact imply a net enhancement
in the proton decay rate, as well as of the $\mu^+K^o$ mode\cite{BPW2}. 
There are of course uncertainties in the prediction for proton-decay rate
owing to uncertainties 
in the SUSY-spectrum, the hadronic matrix element and the
relative phases of the many different contributions (see  
Ref.
\cite{BPW2} for
details).  However, given that the individual contributions to the 
amplitude are enhanced by the
neutrino and fermion mass-effects, and that there are several
prominent channels 
(i.e. $\bar{\nu}_{\tau}K^+$, $\bar{\nu}_{\mu}K^+$ and $\mu^+K^0$), 
it seems that it would be hard 
 to
reconcile the ideas of supersymmetric unification described here, if the
proton life-time exceeds about $10^{34}$ yrs 
\cite{63}.
Assuming such
a unification, the prospect for discovery of proton decay at SuperK and/or
at ICARUS thus seems strong.

\pagebreak

To conclude, with neutrino masses and coupling unification revealed,
{\it proton decay remains 
as the missing link}.  Its discovery, with dominance
of $\bar{\nu}K^+$ and prominence of $\mu^+K^o$ modes, would in fact be a
double confirmation of both 
supersymmetric
unification through G(224)/SO(10), as well as of the ideas of neutrino
masses, described above in this context.  

\vspace{.75in}

{\bf Acknowledgement:}  I have greatly benefitted from discussions on
 topics in this manuscript with Keith Dienes, Alon Faraggi, Qaisar Shafi, 
Joseph Sucher
and
Edward Witten, and especially so, 
from collaborative discussions with Kaladi
S. Babu and Frank Wilczek.  
The research was supported in part by NSF
Grant No. Phys-9119745 and in part by the Distinguished Research
Fellowhsip awarded by the University of Maryland.


\begin{thebibliography}{999}
\bibitem{SIR} Source: "Ideals and Realities" - Selected Essays by Abdus
Salam, Ed. by C.H. Lai, Publ. by World Scientific.
\bibitem{SF} J.C. Pati in Salam Festschrift, Ed. by A. Ali, J. Ellis and
S. Randjbar-Daemi, Publ. by World Scientific, pages (368-391), 1993.
\bibitem{Pati} J.C. Pati, {\it "Recollections of Abdus Salam: Scientist and
Humanitarian"} - 
Physics Today (abridged version), August (1997); News
Letter of the American Chapter of the Indian Physics Association (1997).
\bibitem{SU(2)} S. L. Glashow, {\it Nucl. Phys.} {\bf 22} (1961) 57a;
S. Weinberg, {\it Phys. Rev. Lett.} {\bf 19} (1967)
1269; Abdus Salam, in Elementary Particle Theory, Nobel Symposium, ed.
by N. Svartholm (Almqvist, Stockholm, 1968), p. 367.
\bibitem{GM} For attempts to derive the current algebra framework using
the singlet U(1) vector gluon see e.g. H. Fritzsch and M. Gell-Mann,
Proc. 15th High Energy Conf., Batavia (1972), pages 135-165,
and references there
in.  The difficulty of SU(9) or SU(12) global symmetry associated with
the U(1) vector gluon coupling was one reason for abandoning this
possibility: See J.C. Pati and Abdus Salam, ICTP preprint IC/73/81
(unpublished), and L.B. Okun and V.I. Zacharov, Phys. Lett. {\bf 47B},
258 (1973).
\bibitem{OWG} O.W. Greenberg, Phys. Rev. Lett. {\bf 13}, 598 (1964).
\bibitem{HanNam} M. Han and Y. Nambu, Phys. Rev. {\bf 139}, B1006
(1965).
\bibitem{JPAS} J.C. Pati and Abdus Salam; Proc. 15th High Energy
Conference, Batavia, reported by J.D. Bjorken, Vol. 2, p. 301 (1972);
Phys. Rev. 8 (1973) 1240.
\bibitem{JCPAS} J.C. Pati and Abdus Salam, Phys. Rev. Lett. {\bf 31}
(1973) 661; Phys. Rev. {\bf D10} (1974) 275.
\bibitem{HNSU3} The important suggestion of generating strong
interactions by gauging SU(3)-color symmetry was first made by Han and
Nambu, as early as 1965 \cite{HanNam}.  There were two
shortcomings in this paper which needed to be removed.  First, owing 
to the coupling 
of the photon to integer charge-quarks, $SU(3)^c$
was violated explicitly - rather than spontaneously - a
feature that spoiled renormalizability.  Second, {\it fundamental} strong
interactions were assumed to have two components: the very strong
$SU(3)^c$ gauge force and, in addition, a medium strong force, generated
by the gauge bosons of flavor symmetry, which were identified with ($\rho,\omega,
K^*$, and $\phi$) vector bosons.  These latter were assumed to be 
elementary as much as the octet of color gluons.  However, generating the
electroweak forces via the flavor gauge symmetry, as in SU(2) $\times$
U(1), precludes ($\rho, \omega, K^*$) being 
fundamental.  This in turn uniquely selects out the SU(3)-color gauge 
force as the {\it only source} of fundamental strong interactions 
\cite{JPAS}, 
assuming that all three forces have a gauge-origin.
\bibitem{FGM}  At the time of our suggestion of the combined symmetry at
the 1972 Batavia Conference [Ref. 8], Fritzsch and Gell-Mann also
noted the possibility of
 $SU(3)^c$ gauging at the same conference [Ref. 5].  
They, and many
others, favored the alternative of fractionally charged quarks (fcq) and
unbroken SU(3)-color, with permanent confinement.  With little
understanding of confinement in those days, we thought that
the case of fcq and permanent confinement  should be regarded 
only as an alternative to the case of spontaneously broken
$SU(3)^c$, which would lead to gauge integer charge quarks (icq), with
possibly 
semi-confined but ultimately liberated quarks.  The two alternatives
arise within the same theory depending upon the nature of SSB.  Salam and I, 
(Phys. Rev. Lett. 
{\bf 36}, 11 (1976)), 
as
well as others, notably G. Rajasekaran and P. Roy 
(Pramana {\bf 6}, 303 (1975)),
 therefore proposed
key experiments which could distinguish between the two alternatives.
Eventually, by the 1980's, on the basis of these tests, especially the
two-photon experiments (see e.g. R. Godbole, J. Pati, S. Rindani, T. Jayaraman
and G. Rajasekaran, Phys. Lett. {\bf 142}, 91 (1984)), 
fcq was clearly favored
over gauge icq.  The basic ideas of higher unification, as also the
success of the (combined) 
standard model symmetry, are of course independent of the
nature of quark-charges.
\bibitem{GrWzk} D. Gross and F. Wilczek, Phys. Rev. Lett. {\bf 30}, 1343
(1973); D. Politzer, ibid, {\bf 30}, 1346 (1973).
\bibitem{RMJP} R. N. Mohapatra and J. C. Pati, Phys. Rev. 
{\bf D11}, 566
(1974); Phys. Rev. {\bf D11}, 2558 (1975);
 G. Senjanovic and R. N. Mohapatra, Phys. Rev. D{\bf 12}, 1502
(1975).
\bibitem{Kuzmin} V. Kuzmin, Va. Rubakov and M. Shaposhnikov, Phys. Lett {\bf
BM155}, 36 (1985); M. Fukugita and T. Yanagida, Phys. Lett. {\bf B 174}, 
45 (1986); M. A. Luty, Phys. Rev. {\bf D45}, 455 (1992); W. Buchmuller 
and M. Plumacher, hep-ph/9608308.
\bibitem{JCPBL} J.C. Pati, Proc. Seoul Symposium on Elementary Particle
Physics in memory of B.W. Lee, page 481,(1978).
\bibitem{SU16} J.C. Pati, Abdus Salam and J. Strathdee, Il. Nuov. Com.
{\bf 26A}, 77 (1975).
\bibitem{SU5} H. Georgi and S. L. Glashow, Phys. Rev. Lett. {\bf 32}, 438
(1974).
\bibitem{SO10} H. Georgi, in Particles and Fields, ed. by C. Carlson (AIP,
NY, 1975), p. 575; H. Fritzsch and P. Minkowski, Ann. Phys. {\bf 93},
193 (1975).
\bibitem{19} H. Georgi, H. Quinn and S. Weinberg, Phys. Rev. Lett. {\bf
33}, 451 (1974).
\bibitem{E6} F. G\"{u}rsey, P. Ramond and P. Sikivie, Phys. Lett. {\bf B 
60}, 177 (1976).
\bibitem{SK} The SuperKamiokande Collaboration: "Evidence For
Oscillation of Atmospheric Neutrinos", ICRR-Report-422-98-18 (July,
1998)
\bibitem{JPSK} J.C. Pati {\it "Implications of the SuperKamiokande Result on
the Nature of the New Physics"}, UMD-PP-99-01, hep-ph/9807315; To appear
in the Proc. of the Neutrino-98 Conf. held at Takayama, Japan, June,
1998.
\bibitem{String} See e.g. K. R. Dienes and J. March-Russell, hep-th/9604112; K. R.
Dienes, hep-ph/9606467.
\bibitem{SeeSaw} M. Gell-Mann, P. Ramond and R. Slansky, in:  {\it
Supergravity}, eds. F. van Nieuwenhuizen and D. Freedman (Amsterdam,
North Holland, 1979) p. 315; T. Yanagida, in:  {\it Workshop on the
Unified Theory and Baryon Number in the Universe}, eds. O. Sawada and A.
Sugamoto (KEK, Tsukuba) 95 (1979); R. N. Mohapatra and G. Senjanovic,
Phys. Rev. Lett. {\bf 44}, 912 (1980).
\bibitem{LGR} For reviews, see e.g. P. Langacker, talk at Gatlinburg
Conference, June 94, hep-ph 9411247, and P. Langacker and N. Polonsky, Phys. Rev.
{\bf D52}, 3081 (1995), and references there in.
\bibitem{LLUO} For recent work, see P. Langacker and M. Luo, Phys. Rev. {\bf D 44} (1991) 817;
U. Amaldi, W. de Boer and H. Furstenau, Phys. Lett. {\bf B 260} (1991);
J. Ellis, S. Kelley and D. V. Nanopoulos, Phys. Lett. {\bf B 260} (1991)
131; F. Anselmo, L. Cifarelli, A. Peterman and A. Zichichi, Nuov. Cim.
{\bf A 104} (1991) 1817.  
The essential features
pertaining to coupling unification in SUSY GUTS were noted earlier by
 S. Dimopoulos, S. Raby and F. Wilczek,
Phys. Rev. {\bf D24}, 1681 (1981); W. Marciano and G. Senjanovic, Phys. Rev. D
\underline{25}, 3092 (1982); M. Einhorn and D.R.T. Jones, Nucl. Phys. B
\underline{196}, 475 (1982).
\bibitem{Superstring} M. Green and J.H. Schwarz, Phys. Lett. {\bf 149B}, 117 (1984);
D.J. Gross, J.A. Harvey, E. Martinec and R. Rohm, Phys. Rev. Lett.
{\bf 54}, 502 (1985); P. Candelas, G.T. Horowitz, A. Strominger
and E. Witten, Nucl. Phys. {\bf B 258}, 46 (1985).  For introductions
and reviews, see: M.B. Green, J.H. Schwarz and E. Witten, "Superstring
Theory" Vols. 1 and 2 (Cambridge University Press); M. Dine, ed.,
"String Theory in Four Dimensions" (North Holland, 1988); J. Polchinski,
"What is String Theory?", 1994 Les Houches Lectures, hep-th/9411028.
\bibitem{MTheory} The literature on string-dualities and M-theory is
large.  A few pioneering papers relevant to gauge-coupling unification
are: E. Witten, Nucl. Phys. {\bf B 443}, 85 (1995), P. Horava and E.
Witten, Nucl. Phys. {\bf B 460}, 506 (1996).  For reviews, see e.g. J.
Polchinski, hep-th/9511157; A. Sen, hep-th/9802051, and references therein.
For a layman's guide, see M. Duff (talk at this meeting), hep-ph/9805177 V3.
\bibitem{Ginsparg} P. Ginsparg, Phys. Lett. {\bf B 197}, 139 (1987); V.S.
Kaplunovsky, Nucl. Phys. {\bf B 307}, 145 (1988); Erratum: {\it ibid.}
{\bf B 382}, 436 (1992).
\bibitem{WRev} For an early review on this issue, see e.g. S. Weinberg, Summary
talk, Proc. 26th Intl. Conf. on High Energy Physics, Dallas, Texas
(1992)
\bibitem{Diennes} For a recent discussion, see K. Dienes, Phys. Reports
{\bf 287}, 447 (1997), (hep-th/9602045), and references therein.
\bibitem{Witten} E. Witten, hep-th/9602070.
\bibitem{StGUT} See e.g. D. Lewellen, Nucl. Phys. {\bf B 337}, 61 (1990); A. Font,
    L. Ibanez and F. Quevedo, Nucl. Phys. {\bf B 345}, 389 (1990);
    S. Chaudhari, G. Hockney and J. Lykken, Nucl. Phys. {\bf B 456},
    89 (1995) and hep-th/9510241;
    G. Aldazabal, A. Font, L. Ibanez and A. Uranga, Nucl. Phys.
    {\bf B 452}, 3 (1995); ibid.
    {\bf B 465}, 34 (1996); D. Finnell, Phys. Rev. {\bf D 53},
    5781 (1996); A.A. Maslikov, I. Naumov and G.G. Volkov, Int.
    J. Mod. Phys. {\bf A 11}, 1117 (1996); J. Erler, hep-th/9602032
    and G. Cleaver, hep-th/9604183; and Z. Kakushadze and S.H. Tye,
    hep-th/9605221, and hep-th/9609027; Z. Kakushadze et al, hep-
ph/9705202.  
\bibitem{SMLike} See e.g. K. Dienes and A. Faraggi, Nucl. Phys. {\bf
457}, 409 (1995); C. Bachas, C. Fabre and T. Yanagida, Phys. Lett. {\bf
B370}, 49 (1996).
\bibitem{Babu} J.C. Pati and K.S. Babu, "{\it The Problems of Unification-
Mismatch and Low $\alpha_3$: A Solution with Light Vector-Like Matter}",
hep-ph/9606215, Physics Lett. {\bf B384}, 140 (1996).
\bibitem{JMR} C. Kolda and J. March-Russell (hep-ph/9609480).
\bibitem{JPOak} J.C. Pati, "{\it Baryon Non-Conservation in Unified Theories in
the Light of Supersymmetry and Superstrings}", hep-ph/9611371, Proc.
Oakridge International Workshop on Baryon Instability, March (1996),
pages 7-59.
\bibitem{SOak} See e.g. J. Stone, Proc. Oakridge International Workshop
as in Ref. 37.
\bibitem{SKP} The SuperKamiokande Collaboration: "Search for Proton
Decay Via $p \rightarrow e^+\pi^o$ in a large Water Cherenkov Detector"
ICRR Report - 419-98-15, June 1998.
\bibitem{WS} S. Weinberg, Phys. Rev. {\bf D26}, 287 (1982); 
N. Sakai and T. Yanagida, Nucl. Phys. {\bf B197}, 533 (1982).
\bibitem{Hisano} See e.g. J. Hisano, H. Murayama and T. Yanagida, Nucl.
Phys. {\bf B402}, 46 91993); P. Nath, R. Arnowitt and A.H. Chamseddine,
Phys. Rev. {\bf D32}, 2348 (1985); P. Nath and R. Arnowitt, hep-
ph/9708469 and references therin, and K. Babu (private communications).
\bibitem{Takita} For recent limit on the $\bar{\nu}K^+$ modes, 
see M. Takita,
(SuperK Collaboration), Talk presented in the 29th International
Conference on HE physics, Vancouver, July 1998.
\bibitem{UUDE} See also recent discussion on the importance of the $\Ubar \Ubar
\Dbar \Ebar$ operator, due to Higgsino-dressing (for $\mu \geq 500$ GeV), in 
T. Goto and T. Nihei, KEK-TH-583 (Aug, 1998), and in V. Lucas and S. Raby, 
Phys. Rev. {\bf D55}, 6986 (1997).
\bibitem{Dimop} S. Dimopoulos and H. Georgi, Nucl. Phys. {\bf B 193},
150 91981); N. Sakai, Z. Phys. {\bf C 11}, 153 (1982).
\bibitem{Missing} A. Masiero, D.V. Nanopoulos, K. Tamvakis and T. Yanagida,
    Phys. Lett. {\bf B 115}, 380 (1982); B. Grinstein, Nucl. Phys.
    {\bf B 206}, 387 (1982).
\bibitem{DW} S. Dimopoulos and F. Wilczek, in Proc. Erice Summer School
    (ed. by A. Zichichi), I.T.P. preprint NSF-ITP-82-07 (1981).
    For  recent discussions see K. S. Babu and S. M. Barr, Phys.
    Rev. {\bf D 48}, 5354 (1993);ibid {\bf D 51}, 2463 (1995); 
    S. M. Barr and S. Raby, Phys. Rev. Lett. {\bf 79}, 4748 (1997);
   C. Albright and S. Barr, hep-ph/9712488; Z. Chacko and R. N. 
  Mohapatra, hep-ph/9810315. 
\bibitem{Faraggi} A. Faraggi, Phys. Lett. {\bf B278}, 131 (1992); Phys.
Lett. {\bf B274}, 47 (1992); Nucl. Phys. {\bf B403}, 101 (1993); A.
Faraggi and E. Halyo, Nucl. Phys. {\bf B416}, 63 (1994).
\bibitem{Anto} I. Antoniadis,  G. Leontaris and J. Rizos, Phys. Lett {\bf
B245}, 161 (1990); G. Leontaris, Phys. Lett. {\bf B372}, 212 (1996).
\bibitem{Flipped} I. Antoniadis,
J. Ellis, J. Hagelin and D.V. Nanopoulos, Phys. Lett. {\bf B231},
65(1989).  For a review, see J.L. Lopez and D.V. Nanopoulos,
hep-ph/9511266)).
\bibitem{StProton} J.C. Pati, "{\it The Essential Role of String Derived
Symmetries in Ensuring Proton Stability and Light Neutrino Masses}",
hep-ph/9607446, Phys. Lett. {\bf B388}, 532 (1996).
\bibitem{Fermionic} H. Kawai, D. Lewellen and S.H. Tye, Phys. Rev. Lett. {\bf 57},
     1832 (1986) and Nucl. Phys. {\bf B 288}, 1 (1987); I. Antoniadis,
     C. Bachas and C. Kounnas, Nucl. Phys. {\bf B 289}, 87 (1987). 
\bibitem{Dine} M. Dine, N. Seiberg and E. Witten, Nucl. Phys. {\bf B289}, 585 
(1987);
J.J. Attick, L. Dixon and A. Sen, Nucl. Phys. B {\bf B 292}, 109 (1987).
\bibitem{SU2} See e.g. A. Faraggi, Nucl. Phys. {\bf B428}, 111 (1994).
\bibitem{Ellis} J. Ellis, A. Faraggi and D. Nanopoulos, hep-th/9709049.
\bibitem{U1} A. Faraggi and J.C. Pati, "{\it A Family Universal Anomalous
U(1) in String Models as the Origin of Supersymmetry Breaking and
Squark-degeneracy}", hep-ph/9712516v3, December 97, Nucl. Phys. B (to
appear).
\bibitem{Dvali}  G. Dvali and A. Pomarol, Phys. Rev. Lett. {\bf 77}, 3728
(1996); P. Binetruy and E. Dudas, Phys. Lett. {\bf B 389}, 503 (1996).
\bibitem{FJP} A. Faraggi and J.C. Pati, "{\it Meeting the Constraint of
Neutrino-Higgsino Mixing in Gravity-Unified Theories}", Phys. Lett. {\bf
B 400}, 314 (1997).
\bibitem{K} K.S. Babu and J.C. Pati, "{\it Towards A Resolution of the 
Supersymmetric CP Problem Through Flavor and Left-Right Symmetries}", 
(to appear).
\bibitem{BPW1} K. S. Babu, J. C. Pati and F. Wilczek, Phys. Lett. {\bf
B423}, 337 (1998), hep-ph/9712307.
\bibitem{BPW2} K.S. Babu, J.C. Pati and F. Wilczek, "{\it Linking Fermion
Masses, Neutrino Oscillations and Proton Decay Via Supersymmetric
Unification}"  (to appear).
\bibitem{MS} S. Mikheyev and A. Smirnov, Nuovo Cim. {\bf 9C}, 17 (1986),
and L. Wolfenstein, Phys. Rev. {\bf D17}, 2369 (1978).  
\bibitem{MS2} For a recent
comprehensive analysis of the solar neutrino data, see J.N. Bahcall,
P.L. Krastev and A. Yu. Smirnov, "Where do we stand with solar neutrino
oscillations?", IAS-preprint (July, 1998).
\bibitem{63} To be more specific, as mentioned in the text, the d=5 proton 
decay amplitude in SO(10) 
receives contributions from the standard as well as the new 
operators, the former being inversely proportional to $M_{eff} =
(\lambda < 45_H >)^2/M_{10}$ (see text and Ref. [59]).  Although, 
one expects 
that $\lambda < 45_H > \sim M_U \sim 2 \times 10^{16}$ GeV, $M_{10}$ 
can plausibly
be one to even two orders of magnitude lower than $M_U$, and thus 
$M_{eff}$ can be 
correspondingly higher than $M_U$.  However, unless 
string/GUT-scale threshold 
corrections to 
$\alpha_3 (m_z)$ are larger than about $10 \%$ (with negative sign),
one would expect that $M_{eff} \leq 10^{18}$ GeV, for the sake
of gauge coupling unification [see Ref.[59]].  
In this case, the standard operators by
themselves, evaluated in the light of the SuperK result, lead to proton
lifetimes $\Gamma(p \rightarrow \bar{\nu}K^+)^{-1}_{std} \leq 10^{31 \pm
2}$ yrs.  Note the shortening of the lifetime (especially the theoretical
upper "limit") which (it turns out) arises in part because of the SuperK 
result.  This suggests that given the standard operators in SO(10), proton 
decay via the $\bar{\nu}K^+$ modes should probably have been seen already; 
and at any rate it ought to be seen in the very near future.  Continuing 
absence of such a decay leading to an improvement in the current limit 
of the corresponding lifetime ($ > 5.5 \times 10^{32}$ yrs) [42] 
by a factor of 2-4 would thus cast serious doubt on the standard operators in 
SO(10) as a viable source of proton decay.  Now, as also remarked in the text, in some non-GUT
string models leading to symmetries like G(224) or G(2113), the familiar
color-triplets get projected out, and thus the standard d=5 operators do
not contribute.  It is worth noting that even in this case the new d=5
operators related to neutrino masses can still survive.
Given the SuperK result, these 
new operators by themselves lead to proton-lifetimes typically 
in the range 
of $10^{32 \pm 2.5}$ yrs, for $m_{\tilde{q}} \approx (1-1.5) TeV, 
(m_{\tilde{W}}/
m_{\tilde{q}}) \approx$ (1/6) (1/2 -2), hadronic matrix element $\approx$
(0.006 $GeV^3$) (1/2 - 2) and the mass of the Higgs 16-plet $\approx 2 
\times 10^{16}$ GeV (1/2 - 2).  There is of course no reason why all the 
uncertainties would contribute only towards extending (rather than
some of them shortening) proton lifetime relative to the "central"
value.  Hence, with contributions from just the neutrino-related d=5 
operators, one expects on most reasonable grounds,  
$\tau_{proton} \leq 10^{34}$ yrs (mentioned in the text).  
\end{thebibliography}
\end{document}